\pdfoutput=1
\begingroup\expandafter\expandafter\expandafter\endgroup
\expandafter\ifx\csname pdfsuppresswarningpagegroup\endcsname\relax
\else
\fi

\documentclass[preprint,review,11pt]{elsarticle}
\usepackage[utf8]{inputenc}
\usepackage[margin=1in]{geometry}

\usepackage[hyphens]{url}
\biboptions{sort&compress, square, comma}
\bibpunct{{\unskip~[}}{]}{,}{n}{}{;} 
\usepackage[breaklinks=true, linkcolor=blue, citecolor=blue, colorlinks=true]{hyperref}
\usepackage{graphicx}
\graphicspath{{./figures/}}

\usepackage{caption,subcaption}
\usepackage[version=4]{mhchem}
\usepackage{booktabs,multirow}
\usepackage{latexsym,amsmath,amssymb}
\usepackage[normalem]{ulem}
\usepackage[textsize=small]{todonotes}  

\usepackage[T1]{fontenc}
\usepackage[english]{babel}
\usepackage{csquotes}
\usepackage{textcomp}

\usepackage[binary-units]{siunitx}
\DeclareSIUnit\atm{atm}
\DeclareSIUnit{\calorie}{cal}

\journal{Combustion and Flame} 

\begin{document}

\begin{frontmatter}

\title{The Impact of Roaming Radicals on the Combustion Properties of Transportation Fuels}

\author[neu]{Richard H. West\corref{cor1}}
\ead{r.west@northeastern.edu}

\author[brown]{C. Franklin Goldsmith}
\ead{franklin{\_}goldsmith@brown.edu}

\address[neu]{Department of Chemical Engineering\\ 
Northeastern University, Boston 02115, USA}
\address[brown]{School of Engineering, Brown University\\
Providence, Rhode Island 02912, USA}

\cortext[cor1]{Corresponding author}


\begin{abstract}

A systematic investigation on the effects of roaming radical reactions on global combustion properties for transportation fuels is presented. 
New software was developed that can automatically discover all the possible roaming pathways within a given chemical kinetic mechanism. This novel approach was applied to two mechanisms taken from the literature, one for heptane and one for butanol. Ignition delay times and laminar flame speeds were computed over a broad range of conditions, while testing varying degrees of roaming. As the degree of roaming is increased, the ignition delays increased, consistent with the hypothesis that roaming decreases the reactivity of the system. The percent increase in the ignition delay is strongly temperature dependent, with the largest effect seen in the negative temperature coefficient regime. Outside of this temperature range, the effect of roaming on global combustion properties is small, on the order of a few percent for ignition delays and less than a percent for flame speeds. The software that was used to create the new mechanisms and test the effects of roaming on combustion properties are freely available, with detailed tutorials that will enable it to be applied to fuels other than heptane and butanol.

\end{abstract}

\begin{keyword}
    chemical kinetics\sep roaming radicals\sep combustion\sep ignition delay\sep laminar flame speeds
\end{keyword}

\end{frontmatter}
\clearpage


\clearpage
\section{Introduction}

Broadly speaking, traditional statistical models for unimolecular decomposition of a closed-shell species may be characterized into one of two classes of reactions: 
homolytic cleavage to form two radicals, and some type of molecular elimination to form two closed-shell products; 
the transition state for the former typically has no maximum along the minimum energy path and must be determined variationally, whereas the transition state for the latter typically occurs at or near a first-order saddle point in potential energy. 
In 2004, a third pathway was discovered: the roaming reaction\cite{Townsend2004a}. 
A roaming reaction is a dynamical process in which unimolecular dissociation begins as homolytic cleavage to produce two radicals, but at long inter-atomic distances at which the covalent bond is effectively broken but the potential is otherwise still attractive, the radicals reorient themselves to facilitate an H-atom transfer, thereby producing two closed-shell products. 
Thus, the radicals roam around the otherwise high-energy, entropically disfavored tight transition state for molecular elimination.

Since the work of Townsend et al,\cite{Townsend2004a} there has been an explosion of interest in roaming reactions\cite{Bowman2006a,Lahankar2006a,Bowman2006b,Gupte2007a,Lahankar2007a,Suits2007a,Lahankar2008a,Suits2008a,Heazlewood2008a,Shepler2008a,Goncharov2008a,Saxena2009a,Harding2010a,Sivaramakrishnan2010a,Kamarchik2010a,Bowman2011b,Herath2011a,Sivaramakrishnan2011b,Sivaramakrishnan2011a,Bowman2011a,Hause2011a,Klippenstein2011a,Sivaramakrishnan2012a,Harding2012a,Tranter2013a,Peukert2013a,Klippenstein2013a,Homayoon2013a,Zhang2014a,Joalland2014a,Mauguiere2014a,Isegawa2014a,Lee2014a,Prozument2014a,Maeda2015a,Annesley2015a,Sivaramakrishnan2015a,Tsai2015a,Houston2016a}.
Most of the initial attention to roaming focused on formaldehyde \cite{Townsend2004a,Bowman2006a,Bowman2006b,Lahankar2006a,Lahankar2007a,Suits2007a,Lahankar2008a,Houston2016a} and acetaldehyde \cite{Gupte2007a,Heazlewood2008a,Shepler2008a,Harding2010b,Sivaramakrishnan2010a,Lee2014a,Sivaramakrishnan2015a}, but more recent studies considered larger carbonyls \cite{Goncharov2008a,Saxena2009a,Peukert2013a,Mauguiere2014a,Tsai2015a}, alkanes \cite{Harding2010a,Sivaramakrishnan2011a,Sivaramakrishnan2012a}, ethers \cite{Sivaramakrishnan2011b,Tranter2013a}, and \ce{NO2}-containing compounds\cite{Hause2011a,Homayoon2013a,Zhang2014a,Isegawa2014a,Prozument2014a,Annesley2015a}. 
Collectively, these experimental and theoretical studies suggest that the roaming mechanism is nearly universal in gas-phase chemical kinetics, and that the branching fraction between roaming and homolytic bond fission typically is between 1-10\% under combustion conditions. The magnitude of the branching fraction is highly dependent on the underlying potential energy surface (PES), particularly the energy of the roaming saddle point relative to the two radical fragments at infinite separation.

Despite the ubiquitous nature of roaming, the implications of the roaming mechanism on global combustion properties have not been thoroughly investigated. Presumably, this absence is due to the computational cost associated with a detailed investigation, and not due to a lack of interest or awareness. A truly quantitative prediction of roaming necessitates multireference theory calculations in the PES around the roaming saddle point, followed by some variant of molecular dynamics; at the time of publication, this process typically requires $10^3 - 10^5$ CPU-hours, depending upon the size of the molecule and complexity of the active space in the CASSCF calculations. Given that there could be hundreds of roaming reactions in transportation fuels, a reaction-specific investigation for transportation fuels is computationally unrealistic. 

The aim of this manuscript is to provide a systematic investigation of the impact of roaming on the ignition delay and laminar flame speed of transportation fuels. To that end, we have selected two literature mechanisms: the n-heptane mechanism (v3.1)\cite{Mehl2011a}, and the butanol mechanism\cite{Sarathy2012a}, of Lawrence Livermore National Laboratory. Given the experimental and computational challenges associated with measuring and/or predicting the roaming branching fractions, treating each bond-fission reaction in these mechanisms individually is not feasible. Instead, we make a global approximation: we assume that each bond-fission reaction in the mechanism has the potential for roaming, and that the resulting branching fraction between bond fission and roaming is the same for all molecules. We consider roaming branching fractions of 0, 1, 5, and 10\%. To perform this test, we have written an interpreter that reads the literature mechanism, finds all the bond fission reactions, automatically determines all the feasible roaming pathways, and generates a new mechanism that includes these new reactions. 

This global approach is a useful first pass to determine whether or not roaming reactions have a significant effect on the combustion properties of transportation fuels under industrially relevant conditions. One outcome of this work will be to suggest which roaming pathways should be investigated in greater detail.

\section{Methods}

\subsection{Automated Discovery of Roaming Pathways in Literature Mechanisms}\label{importer}
The overview is to take each kinetic model, in the \textsc{Chemkin} format in which it was published, and generate a new, modified kinetic model with all the roaming pathways added and bond fission reactions adjusted accordingly, then perform ignition and flame simulations to determine the impact of the roaming pathways. 
To automate this workflow as much as possible (there are over 7000 reactions to check for roaming potential and we end up adding over 300 roaming pathways to the two mechanisms) we make extensive use of the open-source Python software Reaction Mechanism Generator (RMG).\cite{Gao:2016dk,RMGPy}
The steps are described here in further detail:

\subsubsection{Identify species}
The first task is to identify the molecular structure of each of the named species in the \textsc{Chemkin} files.
For example,  the  name {\tt IC4H8OOH-TO2} corresponds to the species with the SMILES string {\tt CC(C)(COO)O[O]}, {\tt IIC4H7Q2-T} is {\tt C[C](COO)COO}, and the molecule with the SMILES {\tt CC(=O)COO} is called both {\tt C3KET21} and {\tt CH3COCH2O2H}. To aid with this task we have developed a tool that uses RMG's understanding of how molecules ought to react, to help a user identify the molecular structures of species in a kinetic model based on how that model says the species react. 

The ``importer tool'' source code \cite{West:wz} and a database of kinetic models from the literature with species identified\cite{CombustionMechanism:2017cz} are freely available online; potential users are invited to contact the authors for assistance. With the help of this tool, we identified the molecular structures of all the species in both the butanol and the n-heptane models. The original butanol mechanism contains 431 species and 2346 reactions; the original heptane mechanism contains 654 species and 4846 reactions.

\subsubsection{Filter reactions} \label{filter}

The next task is to filter the thousands of reactions in the kinetic model and determine which reactions might have alternate roaming pathways. The first filter is based simply on the stoichiometry: 
at first we only consider unimolecular dissociation (one reactant, two products) or its reverse, bimolecular recombination (two reactants, one product), because we are looking for homolytic fission reactions.

\begin{figure}[h!]
\begin{center}
\includegraphics[width=88mm]{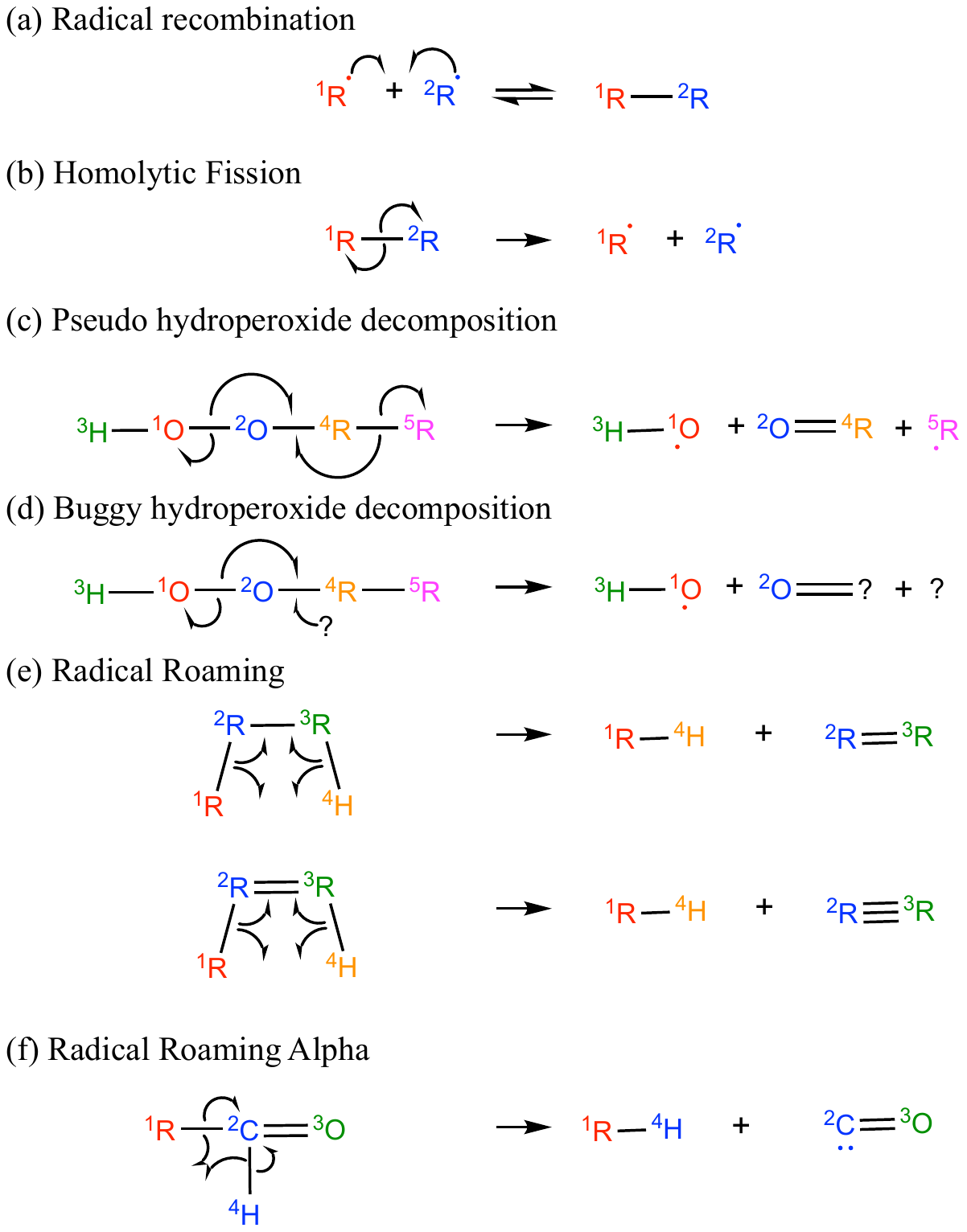}
\caption{Reaction templates and recipes. From \cite{RoamingData2017}}
\label{fig:reaction-recipes}
\end{center}
\end{figure}

The next filter is to see if the reaction can be reproduced using the ``Radical Recombination'' reaction family in RMG. A reaction family in RMG contains a template and a recipe. In this case, the template is simply: two reactants, each with an atom that has one unpaired electron (which will be given labels when the template matches). The recipe says: form a single bond between the two labeled atoms, and decrease the unpaired electron (radical) count on each of them.  This is illustrated in Fig.~\ref{fig:reaction-recipes}a.
The reverse template and recipe, for homolytic fission of a single bond (Fig.~\ref{fig:reaction-recipes}b), is generated automatically by RMG so the reactions can be detected or generated in either direction.
When matching a reaction template, via a subgraph isomorphism algorithm, RMG labels the atoms that are part of the reaction (the superscript numbers in Fig.~\ref{fig:reaction-recipes}), importantly labeling the atoms either side of the bond that is being broken. 
Usually these labels are removed as soon as they have been used to generate an estimate of the reaction kinetics, but we modified the RMG code to preserve these labels, so they could be used to generate roaming pathways in the subsequent step. 

In the butanol model, 185 of the 272 unimolecular dissociation (or bimolecular recombination) reactions fail this reaction family test; in the heptane model 465 of the 667 are filtered out.
Reactions filtered out here are mostly radical addition to a multiple bond, but with some cycloaddition, Diels Alder, and other recombination and addition reaction families.

We then remove any reactions for which the parent molecule (that is about to dissociate) is itself a radical. 
These are typically because RMG properly treats molecular oxygen (which is a triplet in its ground state) as a biradical, and thus \ce{R. + O2 -> RO2.} is seen as radical recombination, but we wish to exclude \ce{O2} roaming and abstracting from the \ce{R.} in the reverse direction. 
This radical filter removes another 5 candidates from the butanol model and 92 from the heptane model.
Finally we check that only one matching pathway is found by RMG, although no reactions were excluded by this check.
This sequence of filters leaves 82 reactions in the butanol model and 110 in the heptane model that each matched a unique radical recombination (homolytic bond fission) reaction in RMG, and are considered candidates for discovering roaming pathways.

However, in some instances, the LLNL models will lump two consecutive elementary steps into a single non-elementary step, and the workflow described above will not be able to catch them. 
An important example of this kind of lumping involves the thermal decomposition of ketohydroperoxides, which are key intermediates in low-temperature auto-ignition. 
Rather than include the resulting keto-alkyloxy radical, the LLNL model automatically has this species undergo beta-scission (typically producing an aldehyde and alkyl or carbonyl radical). 
To find these pseudo-elementary reactions, we filter first for reactions with one reactant and three products (68 in butanol, 111 in heptane), then attempt to match the template for a new RMG reaction family we created called ``Pseudo hydroperoxide decomposition'', illustrated in Fig.~\ref{fig:reaction-recipes}c.
This family finds an additional 18 reactions for the butanol model and 58 for the heptane model that could have alternative roaming pathways.
Finally, we note that some reactions in the original model are quite like the lumped pseudo hydroperoxide decomposition reactions, in that they start with a closed shell hydroperoxide molecule and end with three fragments including an \ce{\cdot OH} radical, but the other two products cannot be generated by applying our reaction recipe.
This may be a deliberate choice to lump isomers and reduce the number of species in the kinetic model, but as we cannot determine a set of elementary steps that reproduce these reactions we term them ``Buggy hydroperoxide decomposition'' (Fig.~\ref{fig:reaction-recipes}d), and include them in the roaming analysis.
There are 16 in the butanol model and 10 in the heptane model.
In total there are 116 reactions for butanol and 178 for heptane that might have alternative roaming pathways.

\subsubsection{Generate roaming reactions}
Not all homolytic bond fission reactions have alternative roaming pathways, and some have more than one. The next step is to generate these alternatives.
For simplicity, we only consider H-atom abstraction for roaming; although the transfer of larger functional groups may be possible, they are neglected.
We created two new RMG reaction families for roaming pathways, each with a template and reaction recipe. 
In the following description \ce{^1R}, \ce{^2R}, and  \ce{^3R} represent any atom.

In the first template (Fig.~\ref{fig:reaction-recipes}e), when there is an H atom beta to the breaking \ce{^1R-^2R} bond, it may be abstracted by the departing radical,
{\em i.e.} in the fission reaction 
\begin{equation}
\ce{^1R-^2R^3RH -> ^1R. + \cdot ^2R^3RH}
\end{equation}
the radical \ce{^1R.} can abstract the \ce{H} from \ce{^3R} leaving a double bond,
giving the overall pathway  
\begin{equation}
\ce{^1R-^2R^3RH -> ^1RH + ^2R=^3R}
\end{equation}
We also allow a double bond to increment to a triple bond, 
{\em e.g.}  
\begin{equation}
\ce{^1R-^2R=^3RH -> ^1RH + ^2R#^3R} 
\end{equation}
This family accounts for most of the roaming pathways. 

In the second family (Fig.~\ref{fig:reaction-recipes}f), created for aldehydes, roaming is possible when there is an H atom alpha to the breaking bond, but only if the fragment is \ce{\cdot CH=O},
{\em i.e.} in the reaction 
\begin{equation}
\ce{^1R-CH=O -> ^1R. + \cdot CH=O}
\end{equation} the departing \ce{^1R.} can abstract the \ce{H} from the \ce{\cdot CH=O} as it leaves,
giving the pathway 
\begin{equation}
\ce{^1R-CH=O -> ^1RH + CO}
\end{equation}
This pathway is included because it is possible (indeed, aldehydes were the first examples of roaming reactions to be discovered), but it is limited to only \ce{CO} as a product because of the unique molecular structure of \"CO  (or \ce{C^+#O^-}); the formation of carbenes through abstraction from an alpha site would be very endothermic.
In the butanol model this creates only one reaction 
\begin{equation}
\ce{CH3CH=O -> CH4 + CO}
\end{equation}
in the heptane model there were 8 instances.

Another modification to RMG was required to make it enforce the atom labels during the subgraph isomorphism detection when matching a species to a reaction template, so that we only generate the roaming reaction that corresponds to our identified homolytic fission reaction (otherwise we'd get all the roaming pathways associated with all the possible homolytic fissions from the parent molecule).
Using the atom labels stored earlier, we can then generate all the roaming pathways corresponding to each identified homolytic fission.
Although the hydroperoxide decomposition templates are asymmetric, the initial elementary step is a symmetrical homolytic cleavage, so in all cases we seek roaming reactions before and after swapping the `1' and `2' labels. 

In total, 120 roaming reactions were generated from 96 parent fission reactions in the butanol model, 
and 198 roaming reactions generated from 163 parent reactions in the heptane model.

\subsubsection{Generate new species properties}\label{new_species}
Some of the new roaming pathways generate species that were not previously included in the kinetic models. For the butanol mechanism, 30 new species were required; for heptane, 56. 
To ensure that the model remain thermodynamically consistent, with all reactions treated reversibly, we  generated thermochemical data for each of these new species. For a few species (isobutyric acid, butanoic acid for the butanol model; prop-2-yn-1-ol, propan-2-ol, pentane-2,4-dione, for the heptane model) RMG found thermochemistry data in a database\cite{Yaws:2012to};
buta-1,2-diene (for heptane) came from USC Mech II\cite{wang2007usc}; the remaining 80 species were estimated using Benson group additivity, as implemented in RMG. Additionally, as required by the flame simulations, we provided transport data (e.g. Lennard-Jones parameters $\epsilon$ and $\sigma$). We used the GRI-Mech 3.0 \cite{GRI3.0} values for butatriene and but-1-yne (for the butanol model) and estimated the rest via critical point properties using the Joback method\cite{Joback:1984vt} in RMG.

\subsection{Modify the rate coefficients}
In all cases, the reaction kinetics in the original models were specified in the bond fission direction, which greatly simplifies the modification to add alternative reaction pathways. The overall rate of decomposition is not changed, but it is assumed to branch into the bond fission and roaming channels.
As a first-order approximation, the temperature dependence of the roaming branching fraction is neglected; instead, only the pre-expontential factor of the original bond-fission reaction is modified. 
The kinetics were all specified in a form which allowed simple modification of the $A$-factor 
(for butanol: 63 Arrhenius, 2 Pressure-dependent Arrhenius ``PLOG'', and 31 Falloff, eg. Troe; for heptane: 151 Arrhenius and 12 Falloff).

For a given roaming fraction $\alpha$, the $A$-factor for the original bond fission channel becomes $A \left( 1 - \alpha\right)$, and the cumulative rate constant for roaming becomes $A \alpha$. 
Thus, the total rate of decomposition is constant, but branching fractions vary with $\alpha$. 
For many bond-fission reactions, more than one roaming product channel is possible. 
In those instances, the individual $A$-factors are further reduced by the number of possible channels. 
For example, in n-butanol, cleavage of the C-C bond between carbons 3 and 4, 
\begin{equation}
\ce{CH3CH2CH2CH2OH -> CH3. + \cdot CH2CH2CH2OH},
\end{equation}
 with the high-pressure limit $A$-factor of $A=3.79\times 10^{24}$ s$^{-1}$ in the original model,
has one roaming pathway, 
\begin{equation}
\ce{CH3CH2CH2CH2OH -> CH4 + CH2=CHCH2OH},
\end{equation}
 so the $A$-factor for that pathway is $A=3.79\times 10^{24} \times \alpha$ s$^{-1}$. In contrast, cleavage of the C-C bond between carbons 2 and 3,
 \begin{equation}
 \ce{CH3CH2CH2CH2OH -> CH3CH2. + \cdot CH2CH2OH}
 \end{equation}
  with $A=5.53\times 10^{24}$ s$^{-1}$ in the original model,
 has two roaming pathways, 
\begin{align}
\ce{CH3CH2CH2CH2OH}&\ce{-> CH2=CH2 + CH3CH2OH}\textrm{ and}\\
\ce{CH3CH2CH2CH2OH}&\ce{-> CH3CH3 + CH2=CHOH},
\end{align}
   so the high-pressure limit $A$-factors for those pathways are each $A=5.53\times 10^{24} \times \alpha/2$ s$^{-1}$.

We repeated calculations with values of $\alpha \in \left\{0.0, 0.01, 0.05, 0.10 \right\}$ (with some additional calculations with an unrealistically high  $\alpha = 0.5$ and 0.9 to ensure the simulations were functioning as expected).

\subsubsection{Generate \textsc{Cantera} input files}
Once our workflow determines all the roaming pathways that are consistent with the original mechanism, it creates a new \textsc{Cantera} input file\cite{Cantera}.
We do this by modifying the \textsc{Chemkin}-to-\textsc{Cantera} conversion script {\tt ck2cti}. 
This is an object oriented script, allowing us to modify the objects in memory
after it parses the \textsc{Chemkin} file but before it writes the \textsc{Cantera} file.
We also update the methods that write the \textsc{Cantera} file, to enable a variable $\alpha$ parameter.

Because the \textsc{Cantera} {\it .cti} input files are in fact executable Python code, expressions such as {\tt (2.53e24*(1-ALPHA))} can be used in place of parameter values, with a single {\tt ALPHA = 0.10} definition at the top of the file, thereby ensuring that all roaming systems are treated consistently throughout the mechanism, and allowing a single file to be quickly modified to explore different values of $\alpha$.

The last step is to identify duplicate reactions and mark them as such for the \textsc{Cantera} solver. 
Of the 120 roaming reactions for butanol, 9 were duplicates of reactions already in the model, and 22 new pairs of duplicates were created.
For heptane, 4 of the 198 roaming reactions duplicated existing reactions and 36 new duplicate pairs were created. These are automatically identified and labeled as duplicates in the updated \textsc{Cantera} input file.

The number of reactions at each stage of the filtering and generation process for each model is summarized in Table~\ref{reaction-counts}.
\begin{table}[h!]
\caption{Reaction statistics for the two models.}
\begin{center}
\begin{tabular}{ l r r }
\hline
								&	Butanol	&	Heptane\\
\hline
Reactions in model   					&  2346	& 4846 \\
1 reactant 2 products, or vice versa		& 272	 &  667 \\
Match ``Radical Recombination'' or reverse & 87	& 202 \\
Parent molecule is not a radical	 		&   82	 & 110\\
\hline
1 reactant 3 products				&  68  	&   111 \\
Match ``Pseudo hydroperoxide decomposition'' &  18  &    58 \\
Match  ``Buggy hydroperoxide decomposition''  &  16    &    10 \\
\hline
Total could have roaming 			  		& 116  &   178 \\
Have 1 or more roaming alternative pathways	&   96   & 163 \\
Total roaming pathways					&  120  &198 \\
New species generated by roaming	&30 	&  56  \\
\hline
\end{tabular}
\end{center}
\label{reaction-counts}
\end{table}%

\subsubsection{Run simulations}

For each value of $\alpha$,  ignition delay times and laminar flame speeds were computed. 
These results were compared to the baseline simulation of original literature mechanism, which was checked to give the same results as the modified model with $\alpha = 0.0$. 

Ignition delay times were calculated using a constant volume adiabatic batch reactor. For the purposes of the simulations, the ignition delay is defined as the time of steepest slope in the temperature profile. 
The ignition delays were computed at three different equivalence ratios: $\phi$ = 0.5, 1.0, and 2.0. 
For $n$-heptane, two different pressures were considered, $P$ = 1 and 50 atm, with temperatures between 600 $< T < $ 2000 K. For the four butanol isomers, a single pressure of $P$ = 43~atm was considered, with temperatures between 1000 $< T < $ 1400~K, which was consistent with the original literature model\cite{Sarathy2012a}. The oxidizer/bath is 20\% \ce{O2} and 80\% \ce{N2} for heptane and 20\% \ce{O2} and 80\% \ce{Ar} for butanol.

Similar conditions were used for the laminar flame speeds: 0.5 $ < \phi < $ 2.0 in a 21:79 mix of \ce{O2} and \ce{N2}, $P$ = 1 
atm, and initial temperatures of $T$ = 300, 600, and 900 K, with an initial domain set to 15 mm. Mixture-averaged transport properties were used.
For the butanol simulations, the flame solver convergence criteria were set to: {\it ratio} = 2, {\it slope} = 0.01, and {\it curve} = 0.01,  which resulted in grids of 975--1778 points and numerical  uncertainties in flame speed, estimated by fitting curves to the plots of flame speed vs.\ grid size, of around 0.25\%, peaking at 1.5\% for the extremes in equivalence ratio.
For the heptane simulations, the convergence criteria were set to: {\it ratio} = 3, {\it slope} = 0.05, and {\it curve} = 0.05,  which resulted in  grids of 253--354 points, and uncertainties of around 2--4\% in flame speed.
 



\subsection{Implementation}
All the algorithms, functions, and simulations described above were implemented in Python,  using Jupyter Notebooks\cite{kluyver2016jupyter} to blend source code, commentary, and results. 
When modifications were required to the source code of other packages such as RMG and \textsc{Cantera}, this was implemented by monkey-patching so that this project is self-contained and does not require a user to recompile customized developer versions of these external packages.
We used \textsc{Cantera} 2.3.0\cite{Cantera} and RMG-Py version 2.1.0\cite{RMGPy}. 
The full source code to generate all the kinetic models, simulation results, and plots seen in this paper, are provided online in reference \cite{RoamingData2017}.

\pagebreak[2]
\section{Results}
\subsection{Ignition delay}

\begin{figure}[h!]
\begin{center}
\includegraphics[width=0.325\textwidth]{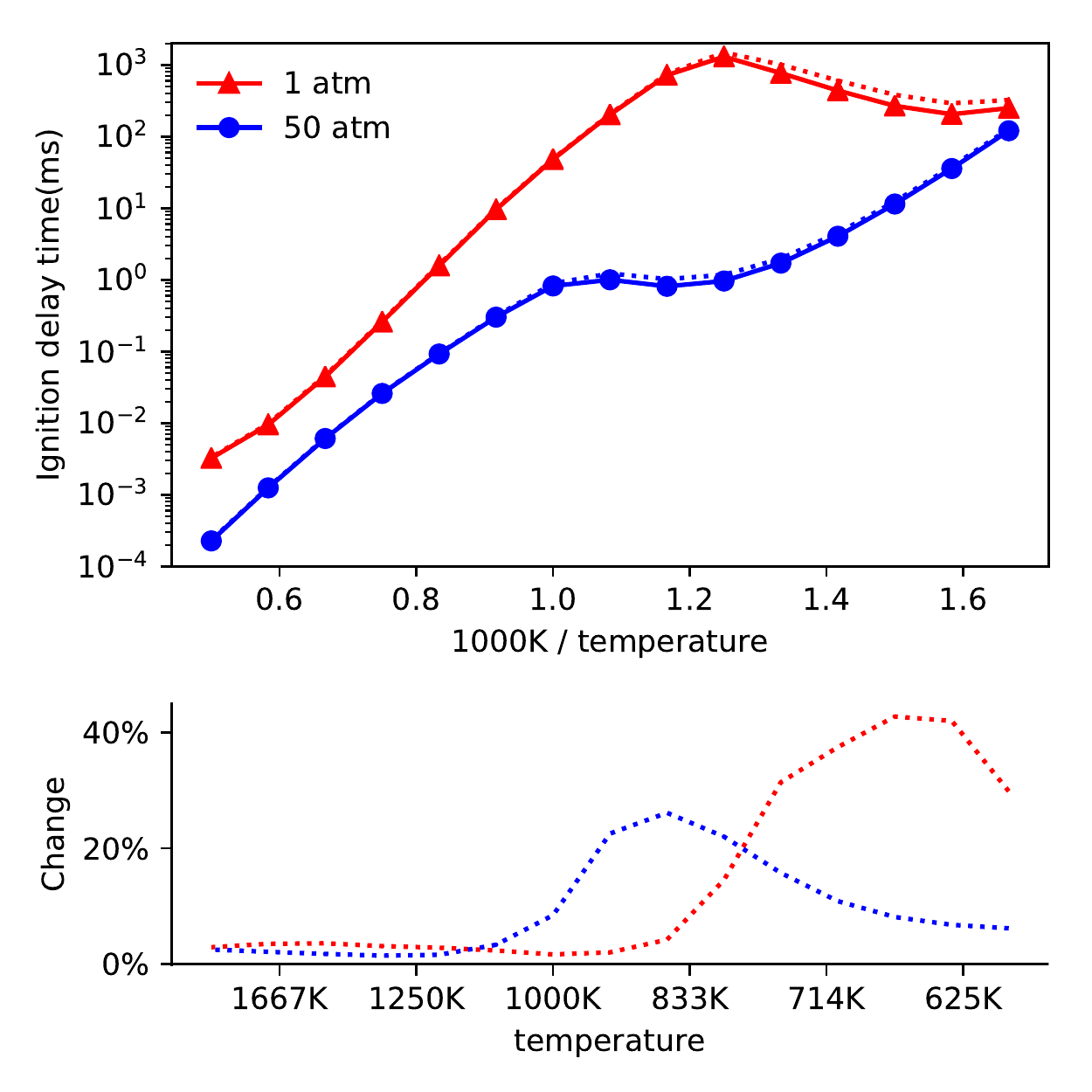}%
\includegraphics[width=0.325\textwidth]{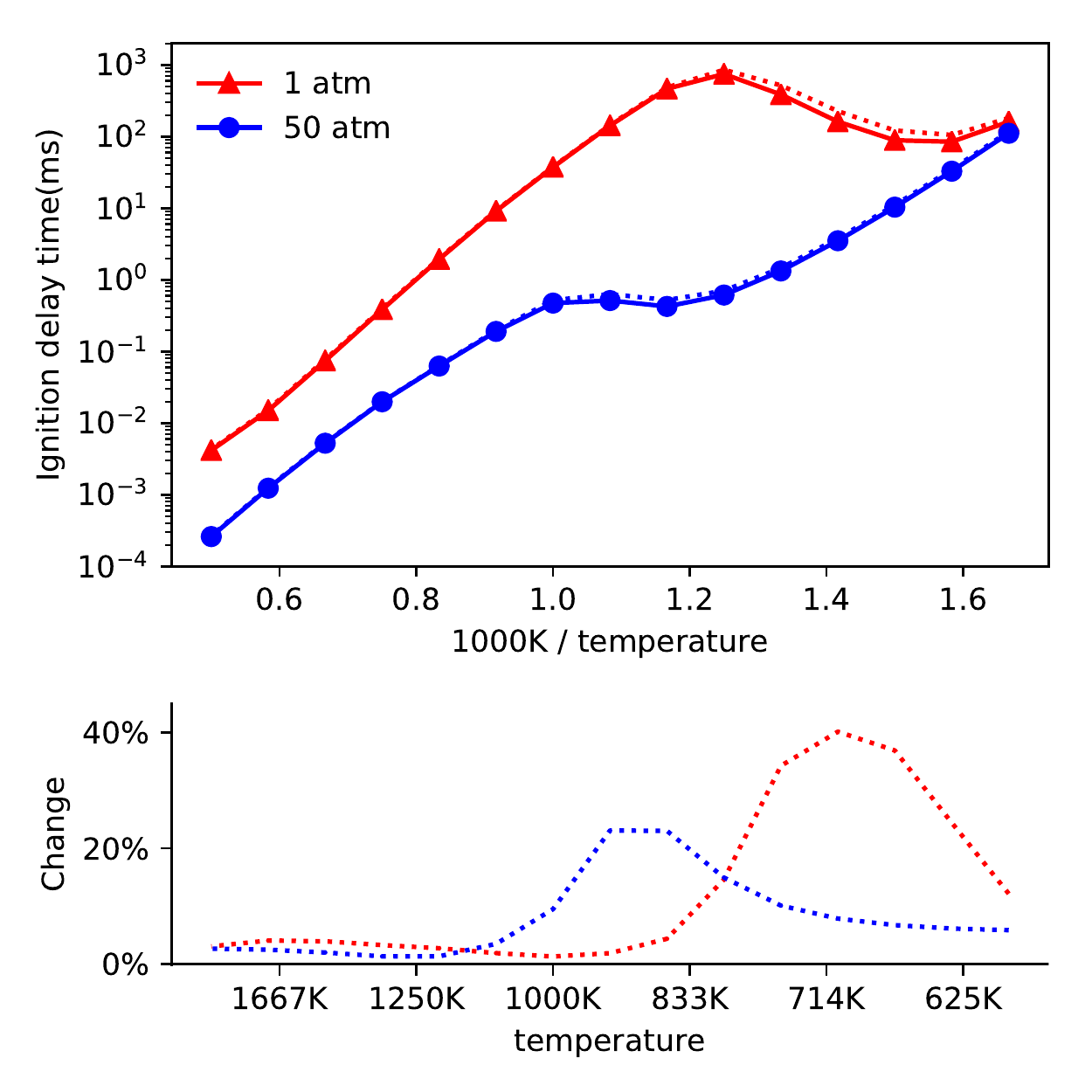}%
\includegraphics[width=0.325\textwidth]{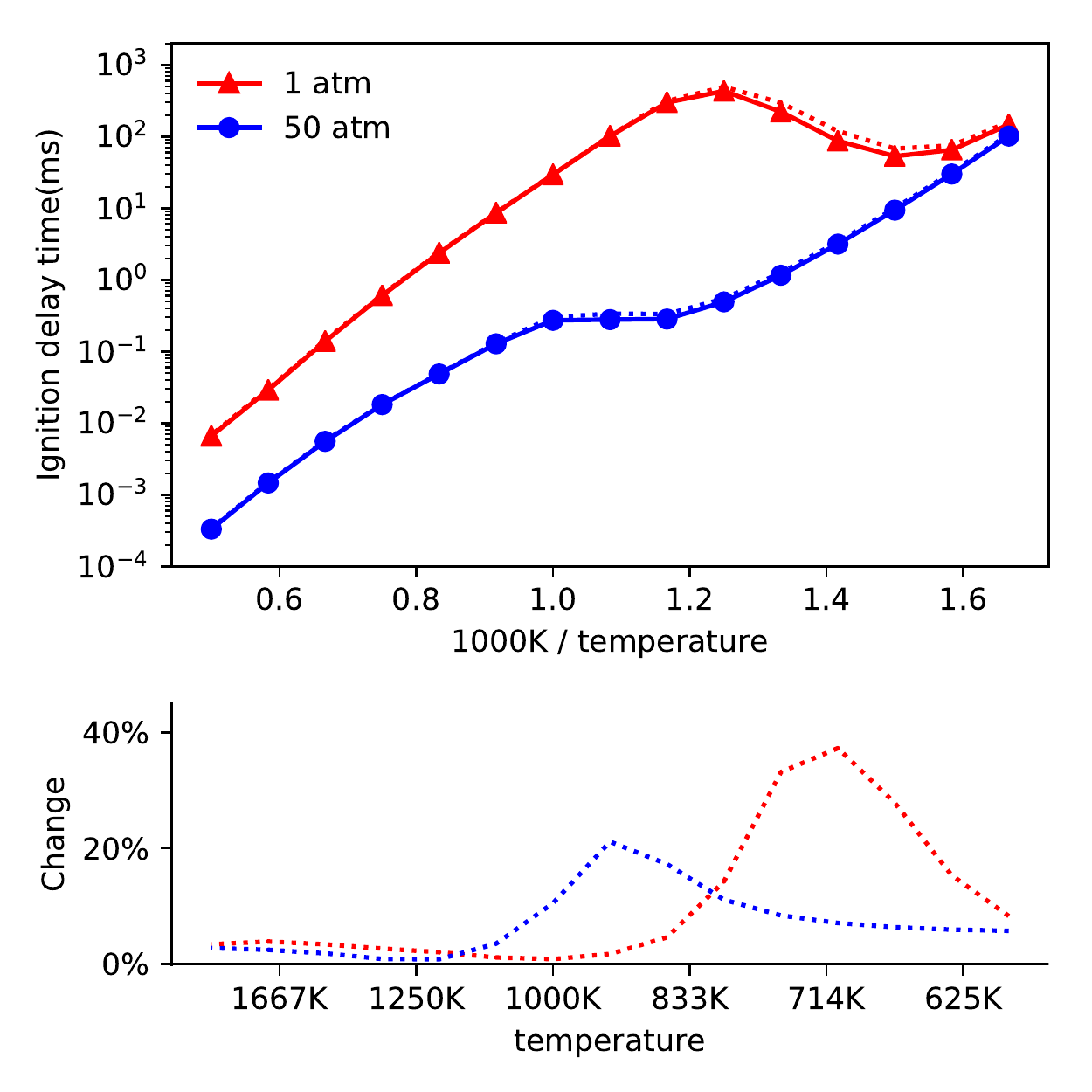}
\caption{Ignition delay times for n-heptane at $P=$1 and 50 atm, $\phi=$0.5 (left), 1.0 (center) and 2.0 (right), in 20\%\ce{O2}/80\%\ce{Ar}. Dotted lines show model with 10\% roaming ($\alpha=0.1$) and solid lines show original model. Lower panel shows percentage change due to roaming. From \cite{RoamingData2017}}
\label{ignition-heptane-10pc}
\end{center}
\end{figure}

The effect of roaming reactions on ignition delays depends strongly on the initial temperature. Fig.~\ref{ignition-heptane-10pc} presents the computed ignition delays for $n$-heptane at equivalence ratios $\phi=0.5$ (left panel), $\phi=1.0$ (center) and $\phi=2.0$ (right)  with a global roaming fraction of $\alpha = 0.10$. 
In both the low-temperature and high-temperature regimes, 10\% roaming increases the ignition delay by less than 5\% at the low- and high-temperature extrema, but it increases the ignition delay by as much as 40\% in the negative temperature coefficient (NTC) regime. This NTC effect also varies with pressure, with lower pressures exhibiting a stronger response to roaming. 

\begin{figure}[htbp]
\begin{center}
\includegraphics[width=0.325\textwidth]{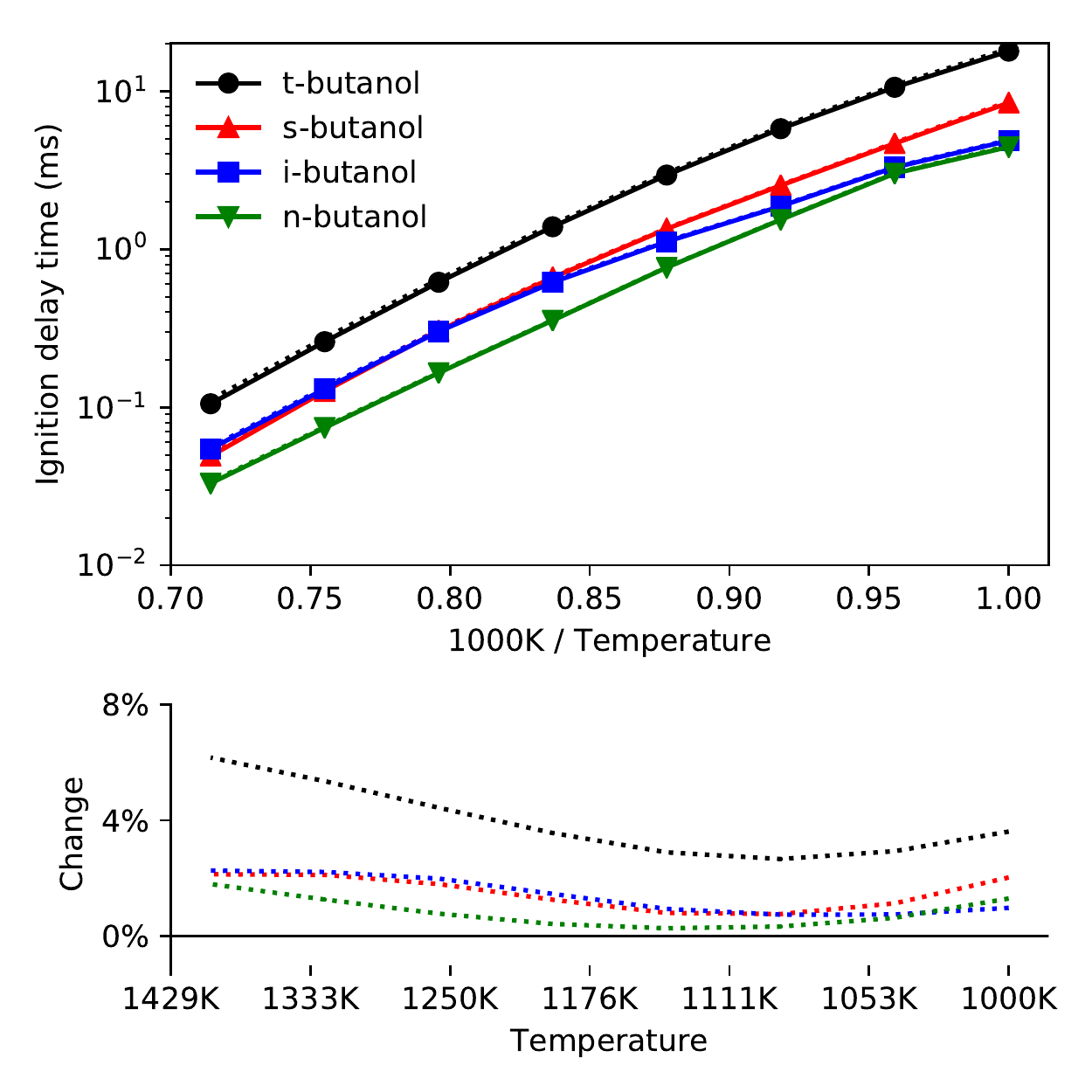}%
\includegraphics[width=0.325\textwidth]{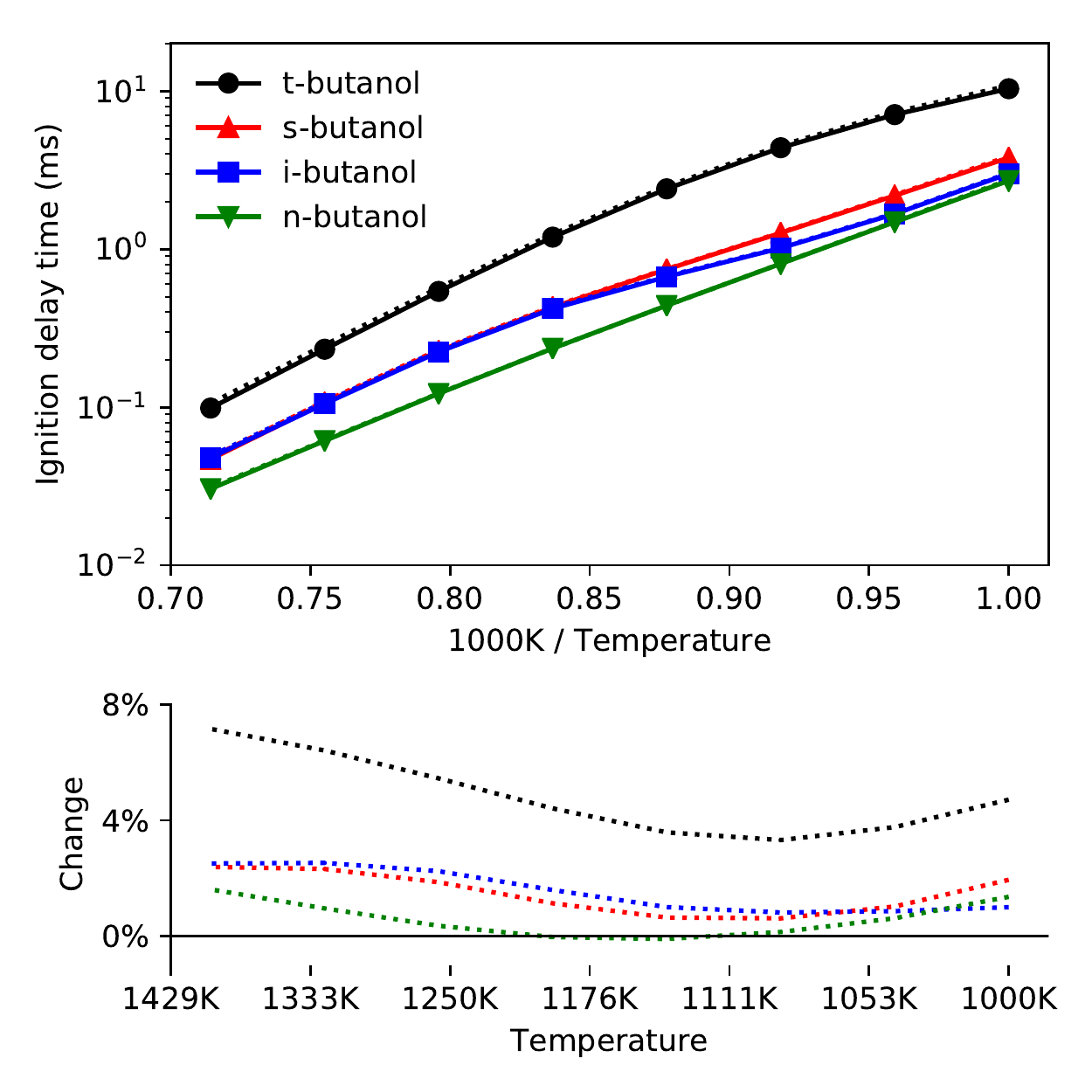}%
\includegraphics[width=0.325\textwidth]{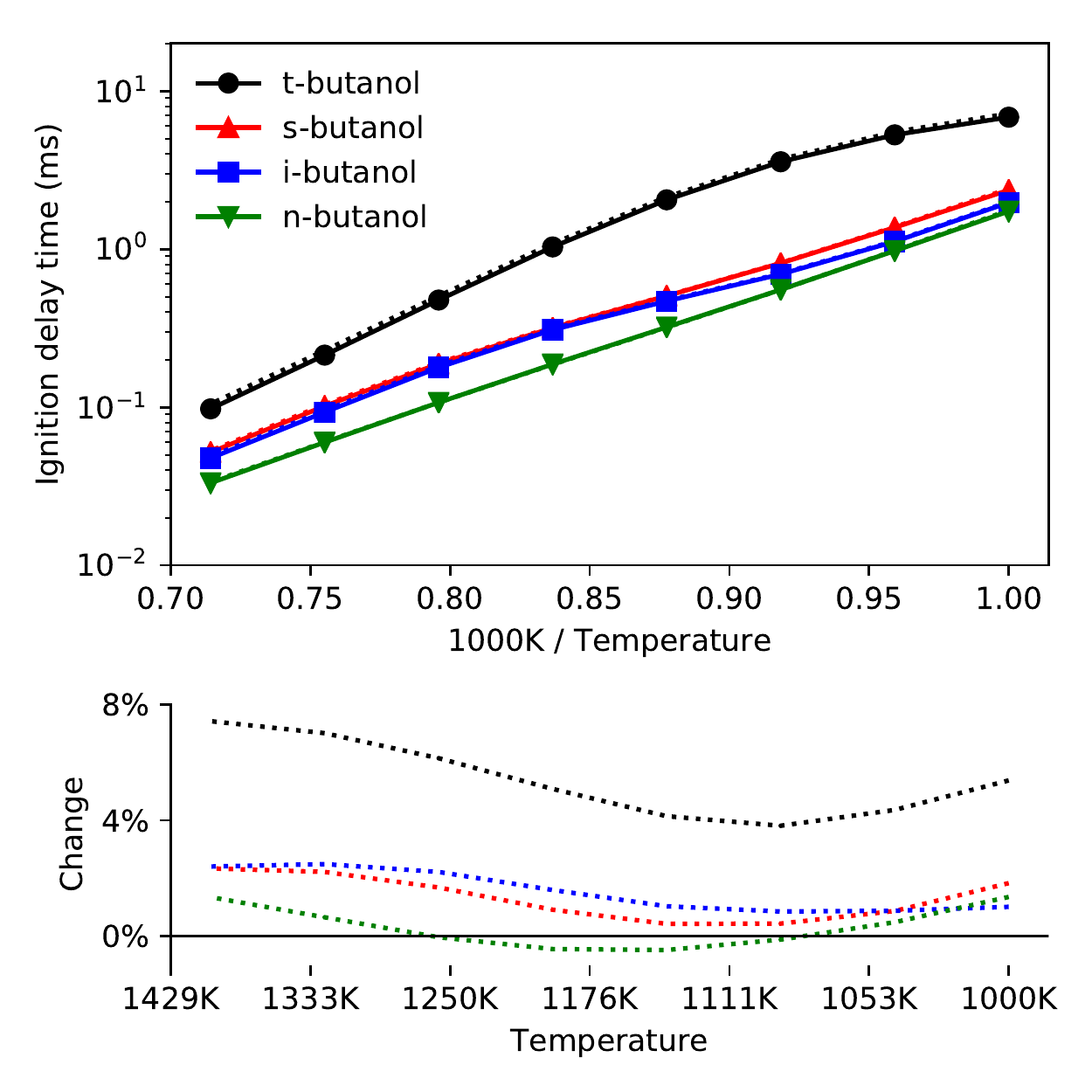}
\caption{Ignition delay times for butanol isomers at $P=43$ atm, $\phi=$ 0.5 (left), 1.0 (center), and 2.0 (right), in 4\%\ce{O2}/96\%\ce{Ar}. Dotted lines show model with 10\% roaming ($\alpha=0.1$). Lower panel shows percentage change. From \cite{RoamingData2017}.}
\label{ignition-butanol-10pc}
\end{center}
\end{figure}

The previous discussion focused on {\em n}-heptane, but the same conclusions hold for the butanol isomers as well (Fig.~\ref{ignition-butanol-10pc}). 
Because there is no NTC region under the conditions of interest, the inclusion of roaming does little to the ignition delays; the effect is most pronounced for {\em t}-butanol, but even there, it is less than 10\%.

\subsection{Laminar flame speed}

\begin{figure}[h!]
\begin{center}
\includegraphics[width=88mm]{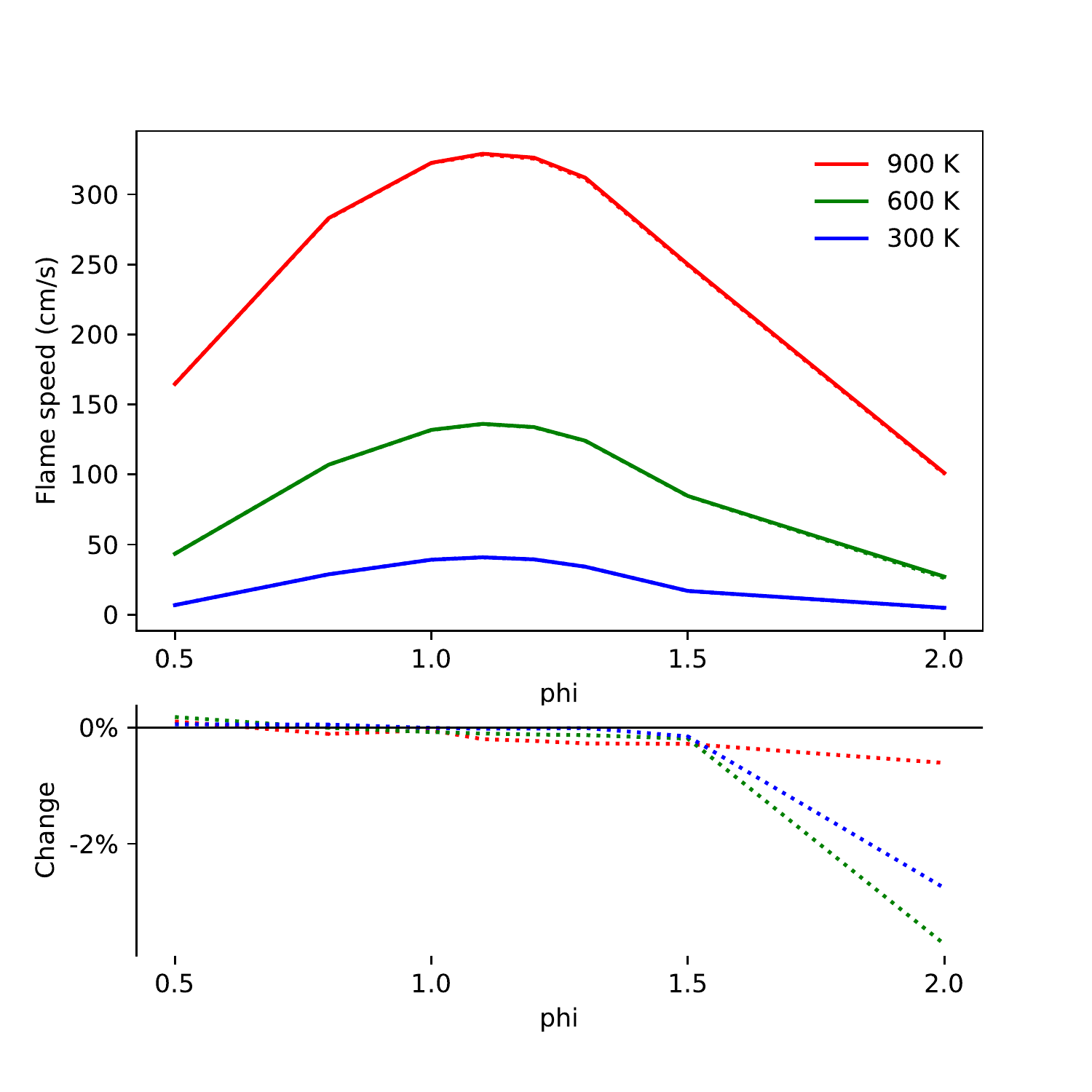}
\caption{Laminar flame speeds for n-heptane  at $P=1$~atm, at $T=$300, 600, and 900~K, $\phi=0.5-2.0$, in 21\%\ce{O2}/79\%\ce{N2}. Dotted lines (mostly hidden by solid lines) show model with 10\% roaming ($\alpha=0.1$). Lower panel shows percentage change. From \cite{RoamingData2017}.}
\label{flamespeed-heptane-10pc}
\end{center}
\end{figure}

\begin{figure}[h!]
\begin{center}
\includegraphics[width=88mm]{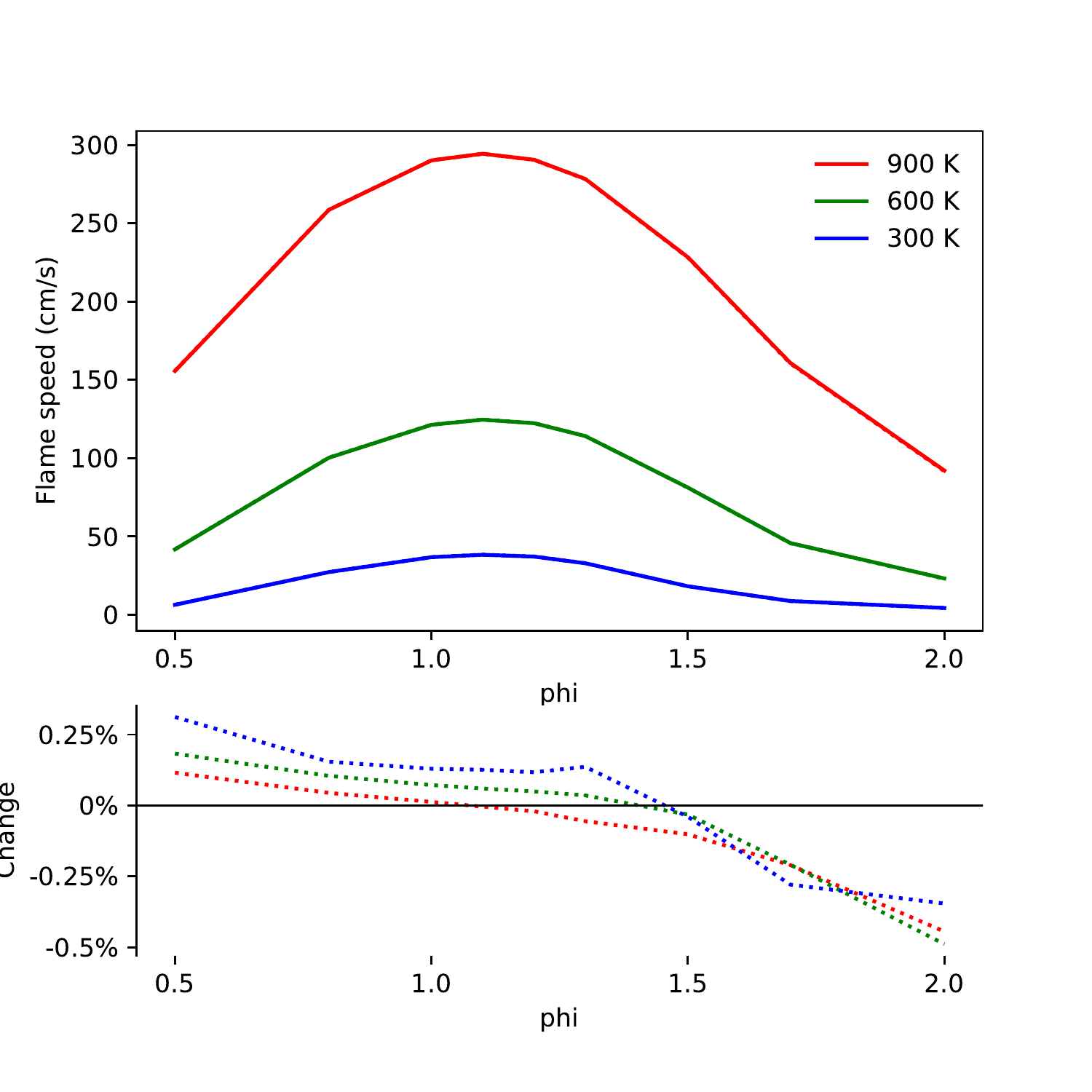}
\caption{Laminar flame speeds for n-butanol  at $P=1$~atm, at $T=$300, 600, and 900~K, $\phi=0.5-2.0$, in 21\%\ce{O2}/79\%\ce{N2}. Dotted lines (mostly hidden by solid lines) show model with 10\% roaming ($\alpha=0.1$). Lower panel shows percentage change. From \cite{RoamingData2017}.}
\label{flamespeed-butanol-10pc}
\end{center}
\end{figure}

The inclusion of roaming has an even more modest impact on the laminar flame speeds, see Fig.~\ref{flamespeed-heptane-10pc} for heptane and Fig.~\ref{flamespeed-butanol-10pc} for butanol.  
The changes for heptane  are within the numerical uncertainty of the simulations (estimated to be 2--4\%).
The butanol simulations were converged to a finer grid, with estimated uncertainties mostly around 0.25\%,
but the effect of roaming is so small -- about  0.25\% -- that it is  still within the numerical uncertainty of the simulations.
That said, butanol seems to exhibit an interesting trend with respect to equivalence ratio, with roaming causing the flame speed to increase under fuel lean conditions and decrease under fuel rich conditions. 


\subsection{Low-pressure burner flames}

Finally, we also calculated the species profiles in low pressure flames.
For n-butanol at 20 torr, premixed with synthetic air, $\phi=1$, $T_0=298$~K,  the maximum variations in mole fractions for species n-butanol, \ce{O2, CO2, H2O, CO}, and \ce{H2} were only about 10--25 parts per million with  $\alpha=0.1$. 
Although the effect is expected to be more pronounced at the extrema in equivalence ratios, changes in major species on the order of 10's ppm is negligible, and those results are not presented in any greater detail.

\section{Caveats}
Our general conclusion is the roaming has a modest effect on combustion properties. 
Having said that, we cannot prove a negative and assert that our conclusion is true for all conditions, and thus there are some caveats that we wish to state.

First, the two mechanisms that we have selected were generated using similar methodologies ({\it e.g.} rate rules) and assumptions. 
We cannot guarantee that other mechanisms generated using alternative assumptions or methodologies would not exhibit greater sensitivity to roaming. 
However, the LLNL mechanisms are highly regarded in the combustion community, and thus they represent a reasonable starting point for analyzing the effects of roaming on the combustion properties of transportation fuels.

Second, we have limited our discussion to ignition delay times and laminar flame speeds. 
We chose these two combustion properties because they are arguably the two most commonly used metrics in mechanism validation for transportation fuels. 
It is possible that other combustion environments, such as a diffusion flame, could exhibit greater sensitivity to roaming. 

Third, as noted in Section \ref{new_species}, a systematic inclusion of roaming required the addition of new species (30 new species to the butanol mechanism and 56 new species to the heptane mechanism). 
However, we did not add subsequent reactions that would couple these products with the other species; they are, kinetically speaking, dead ends. 
Including those ``missing'' reactions is well beyond the scope of this project, since our intention is to test the effect of roaming, not to build a more complete mechanism. Consequently, it is possible that inclusion of secondary chemistry for these new species could have a more dramatic effect on the combustion properties (though, we doubt it).

\section{Discussion}

At first, it is perhaps surprising that modifying all bond-fission (chain-branching) reactions so that 10\% of the flux goes to a roaming (chain-propagating) pathway would have such a modest effect on flame speeds and ignition delays. 
If combustion properties are highly dependent on the growth of the radical pool, and roaming reduces this growth, then surely it should have a large effect. 
Upon closer inspection, however, this result shouldn't be too surprising. From an observational perspective, homolytic cleavage of a single bond rarely turns up in lists of reactions with high sensitivity indices. 
Our goal now is to provide kinetic insight as to why the effect of roaming is most pronounced in the NTC regime but otherwise muted in the low- and high-temperature oxidation regimes.

We begin the analysis with a general rule-of-thumb in chemical kinetics that species/reactions that occur prior to the rate determining step have little effect on the overall rate. 
In high-temperature combustion, the thermal decomposition of the fuel is not the rate determining step, and it generally happens early in the combustion process; accordingly the effect of roaming is, for all intents and purposes, negligible. 

The same conclusion cannot be said for low-temperature combustion, where important bond fission reactions occur after the rate determining step. To see why, it is illustrative to consider the model developed by Merchant et al.\cite{Merchant2015} ~ 
In low-temperature auto-ignition, the first-stage ignition delay can be predicted by considering a small number of reactions that are involved in the second \ce{O2} addition process. 
The penultimate step in this sequence of reactions is the thermal decomposition of a ketohydroperoxide, which as noted in Section \ref{filter}, typically can have a roaming pathway (to produce \ce{H2O} and an aldehyde).  
The sequence of reactions in second \ce{O2} addition form a cycle that is autocatalytic in radical growth: for each OH consumed at the start of the cycle, three OH are produced by the end. 

This cycle can be charactered by a gain, $f_\text{OH}$. The higher the gain, the faster the growth in the radical pool, and thence the shorter the ignition delay. The upper limit of the gain is 3; any loss channels along the cycle will reduce that gain, and if these loss channels are sufficiently high that the gain drops below 1, then the cycle is no longer chain branching. 
Although Ref. \cite{Merchant2015} details several loss channels, the major losses were (i) OH reacting with something other than the fuel to produce the appropriate alkyl radical ($\alpha_\text{OH}$), and (ii) alkylperoxy radicals, \ce{RO2}, that react in some way other than to isomerize to QOOH ($\beta_\text{OH}$). 
Roaming was not considered in their analysis.

At temperatures that are deep into the low-temperature regime, such as 600 K and 50 atm in Figure \ref{ignition-heptane-10pc}, the analytic model suggests that these two loss channels are quite small, and the gain is effectively at its maximum, $f_\text{OH} \approx 3$. 
At these conditions, the rate determining step is the \ce{RO2 -> QOOH} isomerization. 
Unlike the high-temperature analysis, however, the rate determining step precedes the thermal decomposition of the ketohydroperoxide. 
Using the terminology and parameter numbering from Ref. \citenum{Merchant2015},
the ignition delay is sensitive to the thermal decomposition of the ketohydroperoxide, scaling as 
\begin{equation}
\tau \propto \frac{1}{\sqrt{2 k_{14} k_{17}}}, \label{low_T_tau}
\end{equation}
where $k_{14}$ is the rate coefficient for \ce{RO2 -> QOOH}, and $k_{17}$ is the rate coefficient for the homolytic cleavage of the O-O bond in the ketohydroperoxide (details provided in the Supplementary material of Ref. \cite{Merchant2015}). 
According to this simple model, if roaming were to reduce the ketohydroperoxide decomposition channel ($k_{17}$) by 10\%, then the ignition delay would increase by $\sim$ 5\%, since $1/\sqrt{0.9} = 1.054$.
The values at 600 K and 50 atm in Figure \ref{ignition-heptane-10pc}a-c are between 5-8\%, which is in remarkable agreement with the simple model predictions.

As the temperature is increased and we move towards the NTC regime, the behavior is considerably more complicated, and the simple scaling in equation \eqref{low_T_tau}  ceases to be valid. \ce{RO2 -> QOOH} is no longer the kinetic bottleneck, $\alpha_\text{OH}$ and $\beta_\text{OH}$ have deviated from unity, and the shift in the equilibrium constants for \ce{R + O2 <=> RO2} and \ce{O2 + QOOH <=> O2QOOH} represent new loss pathways. The ignition delay time is now highly sensitive to the cumulative effects of these and other losses. Introducing roaming as a new loss at this point further amplifies this sensitivity.

\begin{figure}[htb!]
\begin{center}
\includegraphics[width=88mm]{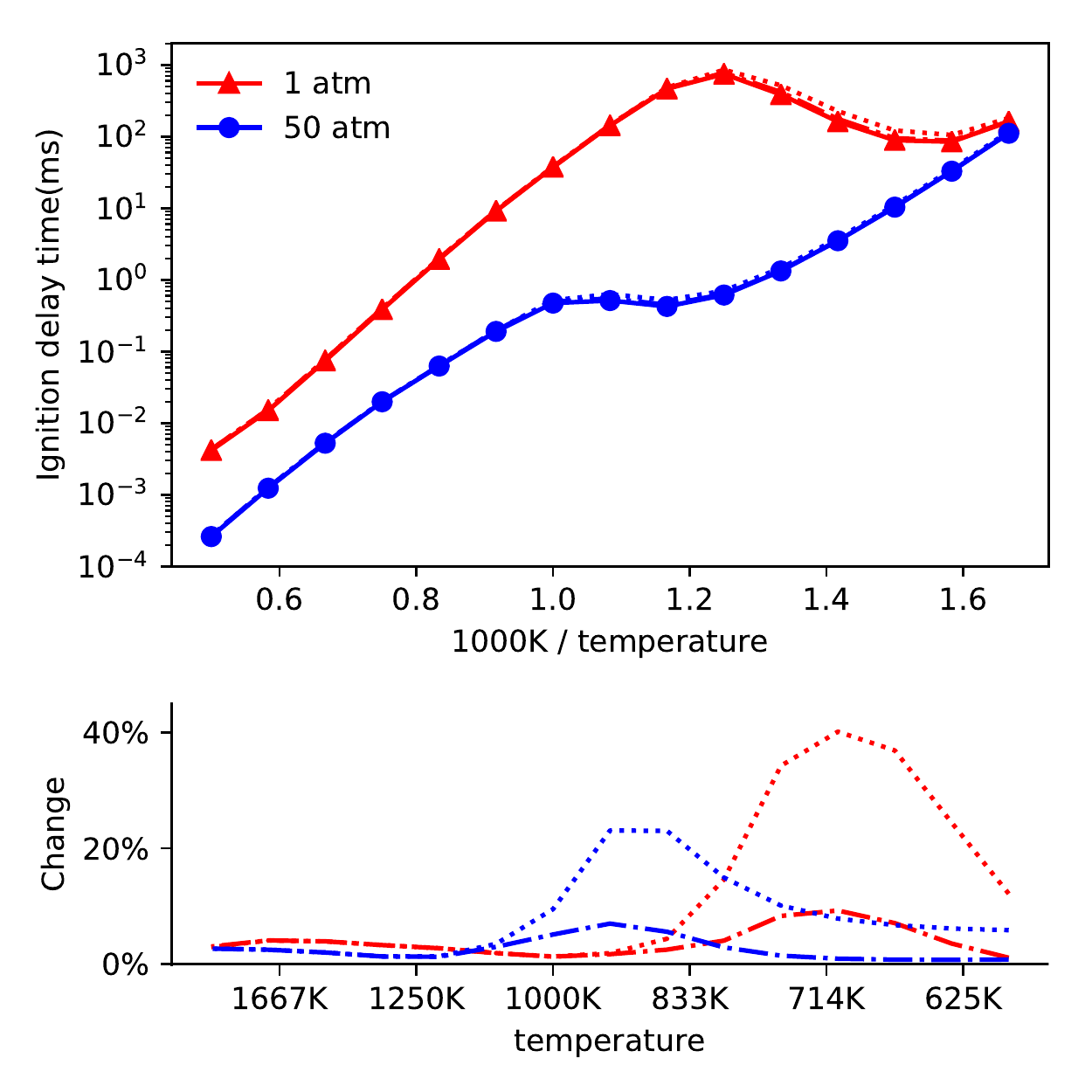}
\caption{Ignition delay times for n-heptane at $P=$1 and 50 atm, $\phi=$ 1.0 in 20\%\ce{O2}/80\%\ce{Ar}. Solid lines show original model, dotted lines show model with 10\% roaming ($\alpha=0.1$), and dash-dot lines show model with 10\% roaming excluding the ketohydroperoxide decomposition channels. Lower panel shows percentage changes due to roaming. From \cite{RoamingData2017}}
\label{ignition-heptane-10pc-threeway}
\end{center}
\end{figure}

Indeed, we can test this hypothesis by turning off the roaming pathways for the ketohydroperoxides that were identified through the ``pseudo hydroperoxide decomposition'' and ``buggy hydroperoxide decomposition'' templates.
 Figure~\ref{ignition-heptane-10pc-threeway} presents ignition delay times for $P=$1 atm and $\phi = 1.0$ when the $\alpha$ for the  hydroperoxide decomposition families are set to zero, but all other roaming reactions remain at $\alpha = 0.1$. 
 Without the possibility of roaming in ketohydroperoxides, the net effect of roaming is substantially reduced in the NTC region. Consequently, we conclude that future efforts on roaming should focus on systematic rate rules for ketohydroperoxides, since these species appear to have the most important influence on ignition delays.

In the case of the flame speeds, the results suggest that the effect of roaming is inconsequential. The simulations exhibit a trend in which the flame speeds increase under fuel lean conditions and decrease under fuel rich conditions. However, the effect is too small to matter, and there is little point in speculating on a chemical cause for this trend when the result is effectively smaller than the error than can be expected for grid convergence in a flame speed simulation. The fact that the effect of roaming is so small is consistent with our interpretation of the high-temperature auto-ignition results: the roaming reactions occur prior to the kinetic bottleneck in the flame, and thus have negligible effect.

All of these conclusions scale with the size of the roaming parameter $\alpha$. If the percent roaming is decreased to 1\% from 10\%, the results are qualitatively similar, but the scale in the percent difference is reduced by approximately an order of magnitude.

This work is part of a broader trend to reexamine fundamental assumptions of chemical kinetics in the context of combustion chemistry. 
For example, each product of a gas-phase reaction is thermalized via a sequence of inelastic collisions with an inert bath gas, and conventional rate theory assumes that thermalization is completed prior to subsequent reactions involving that product. 
Recent work calls into question the validity of this assumption, by noting that rovibrationally excited products can undergo both bimolecular\cite{Bohn1996a,Maranzana2007a,Asatryan2010a,Glowacki2012a,Glowacki2013a,Pfeifle2014a,Burke2015a} and unimolecular\cite{Goldsmith2015a,Labbe2016a,Labbe2017a} reactions prior to collisional thermalization. 
When these ``hot-radical'' reactions are incorporated into chemical kinetic mechanisms for combustion, the overall reactivity of the system increases ({\it i.e.} a consistent decrease in ignition delay times and increase in laminar flame speeds)\cite{Burke2015a,Goldsmith2015a,Labbe2016a,Labbe2017a}.

To a certain extent, roaming radical reactions act as a counter weight to the reactivity-promoting effects of hot radical reactions; by decreasing the rate of radical production, they decrease the overall reactivity of the system. Naively, we might speculate (or hope) that the hot radicals and roaming radicals would cancel each other, but the present work suggests otherwise. Hot radicals appear to have an important effect on flame speeds, whereas the effect of roaming on flame speeds appears to be negligible. 

Together, this growing body of work suggests gas-phase chemistry is far from a solved problem, and that truly predictive models for chemical kinetics in combustion require a fundamental reconsideration of the role of chemical dynamics.

\subsection{Conclusions}
Two literature mechanisms were modified to include roaming radical reactions in a systematic manner. 
To accomplish this goal, new software was developed that could automatically discover all the possible roaming pathways within the given mechanisms. 
The resulting mechanisms were used to compute the ignition delay times and laminar flame speeds for $n$-heptane and four butanol isomers over a broad range of conditions.  
The modeling results suggest that roaming reactions have a modest impact on global combustion properties. 
Even if 10\% of all bond-fission reactions were to end up in a roaming channel, the effect on flame speeds is negligible. The effect on ignition delays is more nuanced, with the most significant effect in the NTC regime. This sensitivity to roaming in the NTC can be understood in terms of the role of thermal decomposition of ketohydroperoxides as a kinetic bottleneck. Future efforts should focus on quantifying the upper limit of roaming in the thermal decomposition of ketohydroperoxides.

\section*{Acknowledgements}

The project was made possible by collaborative tools GitHub, Slack, FaceTime, Screen Sharing, and Intelligentsia. 
CFG gratefully acknowledges support from Brown University, and RHW the same from Northeastern University.
The identification of species in kinetic models is based upon work supported by the National Science Foundation under Grant Nos. 1403171 and 1605568.


\section*{Supporting Information}
As part of a broader trend towards transparency, reproducibility, and sharing in computational engineering, the authors have made all the software available open source to anyone who wishes to use it. 
Furthermore, all the conversion, simulations, and analyses were done using well-documented notebooks in Python. 
These are available on FigShare at \url{https://doi.org/10.6084/m9.figshare.5267365} \cite{RoamingData2017}


\bibliography{bibliography}
\bibliographystyle{elsarticle-num}

\end{document}